\documentclass[fleqn,10pt]{wlscirep}
\usepackage[utf8]{inputenc}
\usepackage[T1]{fontenc}

\usepackage{cite}
\usepackage{amsmath,amssymb,amsfonts}
\usepackage{algorithmic}
\usepackage{graphicx}
\usepackage{textcomp}
\usepackage{xcolor}
\usepackage{braket}  
\usepackage{svg}

\usepackage{tabularx}
\usepackage{adjustbox}
\usepackage{cleveref}
\usepackage{fancyvrb}
\usepackage{setspace}

\usepackage{booktabs}
\usepackage{multirow}
\usepackage{float}
\usepackage{svg}
\usepackage{enumitem}
\usepackage{url}
\usepackage{fancyhdr}
\DeclareUnicodeCharacter{2212}{-}

\def\BibTeX{{\rm B\kern-.05em{\sc i\kern-.025em b}\kern-.08em
    T\kern-.1667em\lower.7ex\hbox{E}\kern-.125emX}}

\setitemize{leftmargin=*}
\setenumerate{leftmargin=*}

\usepackage{titlesec}

\title{Quantum Neural Networks: A Comparative Analysis and Noise Robustness Evaluation}

\author[1,2]{Tasnim Ahmed}
\author[1,2]{Muhammad Kashif}
\author[1,2,*]{Alberto Marchisio}
\author[1,2]{Muhammad Shafique}
\affil[1]{eBrain Lab, Division of Engineering, New York University Abu Dhabi, UAE}
\affil[2]{Center for Quantum and Topological Systems (CQTS), NYUAD Research Institute, New York University Abu Dhabi, UAE}
\affil[*]{Corresponding Author: alberto.marchisio@nyu.edu}

\begin{abstract}
In current noisy intermediate-scale quantum (NISQ) devices, hybrid quantum neural networks (HQNNs) offer a promising solution, combining the strengths of classical machine learning with quantum computing capabilities. However, the performance of these networks can be significantly affected by the quantum noise inherent in NISQ devices. In this paper, we conduct an extensive comparative analysis of various HQNN algorithms, namely Quantum Convolution Neural Network (QCNN), Quanvolutional Neural Network (QuanNN), and Quantum Transfer Learning (QTL), for image classification tasks. We evaluate the performance of each algorithm across quantum circuits with different entangling structures, variations in layer count, and optimal placement in the architecture. Subsequently, we select the highest-performing architectures and assess their robustness against noise influence by introducing quantum gate noise through Phase Flip, Bit Flip, Phase Damping, Amplitude Damping, and the Depolarizing Channel. Our results reveal that the top-performing models exhibit varying resilience to different noise gates. However, in most scenarios, the QuanNN demonstrates greater robustness across various quantum noise channels, consistently outperforming other models. This highlights the importance of tailoring model selection to specific noise environments in NISQ devices. 
\end{abstract}

\begin{document}

\flushbottom
\maketitle

\thispagestyle{empty}


\section{Introduction}
Quantum Computing (QC) is a new computational paradigm that can potentially solve computationally intensive problems substantially faster than its classical counterpart. Over the last few years, there has been significant progress towards the development of quantum computers and the Noisy-Intermediate Scale Quantum (NISQ) devices~\cite{Preskill} have already been developed. These devices have a limited number of qubits and demonstrate limited resilience to noise.
Despite the limitations, the NISQ devices are already being used to demonstrate the so-called quantum advantage for various applications~\cite{google_qc, ibm_qc, qc_adv1}. 
The development of NISQ devices has opened avenues to explore a wide range of applications for the post-quantum era.  
One such application is Quantum Machine Learning (QML)~\cite{Schuld_QML_HS, Schetakis_ReviewQML, Markidis_ProgrammingQNNs,  Massoli_LeapQCQNN}.

QML combines the field of Classical Machine Learning (ML) and QC, and aims to benefit from the development of advanced classical ML models, architectures, training methodologies, and infrastructure to build ML algorithms, such as Deep Neural Networks (DNN) in conjunction with the distinctive computational capabilities of quantum computers~\cite{Huang_2021a, Kubler:2021, Abbas:2021}. 
Inspired by the tremendous success of classical DNNs, the Quantum Neural Networks (QNNs) stand out as one of the most studied QML algorithms~\cite{Schuld:2014}. 
However, due to the limitations of the NISQ devices, the development and practical realization of standalone QML algorithms including QNNs is challenging. 
This has lead to the development of hybrid quantum-classical neural networks (HQNNs), which are designed to be NISQ-compatible and are notably robust against errors typical in NISQ devices.
The fundamental part of an HQNN is the Variational Quantum Circuit (VQC), which utilizes parameterized quantum gates. The parameters of these gates are \emph{trainable} which are optimized during the training using \emph{classical} methods~\cite{pennylane_diff}. 

\begin{figure*}[ht]
    \centering
    \includegraphics[width=\linewidth]{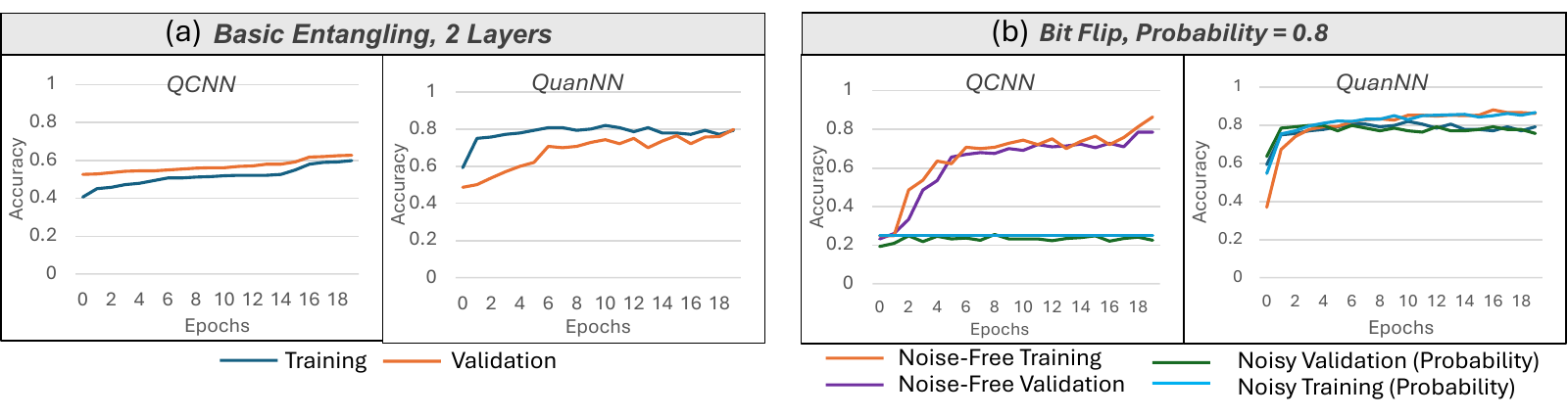}
    \caption{Motivational Case study. (a)Two different variant of HQNN (QCNN and QuanNN) yields different performance with the same underlying architecture of quantum layers (b) Different effects of Bit Flip noise on the performance of QCNN and QuanNN highlight unique noise sensitivities of different HQNNs.}
    \label{fig:motivation}
\end{figure*}


HQNNs have garnered significant interest due to their expressiveness and reduced trainable parameters, made possible by the VQC~\cite{Kashif_HQNN:2022}. These networks have been proposed for diverse applications such as generative modeling~\cite{generative1:2018, generative2:2023}, classification~\cite{farhi2018classification}, and regression analysis~\cite{Regression:2022}, thus prompting the development of various HQNN algorithms. Examples include Quantum Recurrent Neural Networks, Quantum Reinforcement Learning Models, and Quantum Graph Neural Networks, each contributing to the broad QML landscape. This paper examines three frequently used HQNN models used in image classification tasks: Quanvolutional Neural Networks (QuanNN)~\cite{QuanNN:2019}, Quantum Convolutional Neural Networks (QCNN)~\cite{QCNN:2019}, and Quantum Transfer Learning (QTL)~\cite{QTL:2020}, focusing on their unique architecture and performance.

Each of these models uses quantum circuits uniquely to leverage quantum mechanics for machine learning tasks. Both QuanNN and QCNN draw inspiration from classical Convolutional Neural Networks. The QuanNN~\cite{QuanNN:2019} uses a single quantum circuit as a filter which slides over the entire image and extract the features which can then be trained using either quantum circuits or classical layers. 
The QCNN, on the other hand, downsizes the input image to match the first circuit's qubit count and performs pooling by reducing the number of qubits in successive circuits until reaching the desired output size~\cite{QCNN:2019}. A key difference between QuanNN and QCNN is that the QuanNN does not have a pooling operation. 
The QTL model is inspired by classical transfer learning. It involves transferring knowledge from a pre-trained classical network to a quantum setting, where a quantum circuit is integrated for quantum post-processing~\cite{QTL:2020}.

Numerous studies have been conducted to explore the circuit architecture and the trainability challenges for the above-discussed algorithms~\cite{HurQCNN:2022, zaman2024comparative, kashif2024investigating, winderl2023quantum}. However, a comprehensive study that compares the performance of these algorithms for a specific application is still lacking. Additionally, most of the state-of-the-art studies primarily focus on their performance in ideal conditions and do not take into consideration the effect of different quantum noise, which is an inherent characteristic of NISQ devices.

\subsection{Related Work}

The work in~\cite{HurQCNN:2022} investigates the performance of various QCNN models, which differed in their structures of parameterized quantum circuits, quantum data encoding methods, and classical data preprocessing approaches when applied to MNIST datasets. That study reported that QCNNs achieved high classification accuracy despite a limited number of free parameters. However, their analysis was limited to noise-free model performance. Whereas, \textit{our work evaluates the best performing QCNN model under various quantum noise channels, investigating whether or not the model exhibits robustness to noise over time}.

An extensive comparative analysis of different QNN algorithms, including QuanNN, QCNN, and Quantum Residual Network, is presented in~\cite{zaman2024comparative}. The paper examined the effects of varying circuit depths, the number of qubits, and different entanglement settings on image classification tasks using the MNIST dataset. Although the paper provided a thorough evaluation of different models under varying circuit architectures, it did not analyze the performance implications of noise on these models. \textit{In this paper we expand upon this analysis by examining the robustness of the best-performing model under different noise channels. }

The impact of 5 different noise channels on the trainability of HQNNs is investigated in~\cite{kashif2024investigating}. Their study employed a two-qubit VQC within the HQNN architecture, focusing on its performance in binary classification tasks. In contrast, \textit{our paper conducts a similar analysis with the same noise channels but extends it to larger 4-qubit circuits for multiclass classification tasks on MNIST, highlighting the robustness of larger circuits against noise}. 

These studies offered substantial insights into the performance of various models under ideal conditions but provided minimal understanding of the interplay between noise and circuit architecture. Therefore, we believe that a thorough exploration of different noise channels associated with NISQ devices, particularly from the architectural perspective of HQNNs, remains an inadequately addressed area.

\subsection{Motivational Analysis}

In \Cref{fig:motivation}, we underscore the necessity for a thorough comparative analysis of various HQNN architectures and their robustness against different types of quantum noise. 

Our observations indicate that even under the same experimental settings and identical design of the underlying quantum layer, the performance of different HQNN models vary significantly. For instance, the QuanNN model outperforms the QCNN by approximately 30\% in terms of validation accuracy, as depicted in Fig.~\ref{fig:motivation}(a). \textit{This variation in performance highlights the importance of conducting a comprehensive comparative analysis of different HQNN architectures}.

Additionally, the impact of quantum noise on these networks is a critical factor for NISQ devices. As shown in Fig.~\ref{fig:motivation}(b), the same type of quantum noise affects different HQNN architectures, such as QCNN and QuanNN, in distinct ways. \textit{Hence, a comprehensive analysis is needed to understand, how the noise effects the overall performance of HQNNs and what level of robustness different HQNN variants can provide against various quantum noise types}.


\subsection{Our Contributions}
Our key contributions are summarized below: 
\begin{itemize}
\item  \textbf{Comprehensive Comparative Analysis of Different HQNNs.} We conduct a thorough comparison of various HQNN algorithms, including Quantum Convolution Neural Network (QCNN), Quanvolutional Neural Network (QuanNN), and Quantum Transfer Learning (QTL), across a range of circuit architectures. Our analysis spans various entangling structures, variations in layer count, and optimal positioning within the overall network, providing a detailed assessment of each algorithm's effectiveness in multiclass classification tasks under noise-free conditions.

\item \textbf{Evaluation of Noise Robustness.} We select the best-performing HQNN architectures from the previous step and systematically assess the robustness of these algorithms against various types of quantum noise across different probabilities. This involves introducing different quantum gate noise models, such as Phase Flip, Bit Flip, Phase Damping, Amplitude Damping, and the Depolarizing Channel, to determine how different noise sources affect the performance of the selected HQNN architectures.

\item \textbf{Identification of Robust HQNN Architectures.} By evaluating the impact of noise on various HQNN algorithms, we identify architectures that demonstrate resilience to specific quantum noise channels. Our results indicate that QuanNN generally exhibits greater robustness across multiple quantum noise channels, suggesting its potential as a reliable choice for applications in noisy intermediate-scale quantum (NISQ) devices.


\item  \textbf{Guidance for HQNN Design in NISQ Devices.} Our findings provide valuable guidance for designing HQNN architectures in NISQ environments. By demonstrating the varying resilience of different models to specific noise channels, we offer a framework for selecting appropriate architectures based on the noise characteristics of a given quantum device, contributing to the development of more robust and reliable quantum-classical hybrid networks.
\end{itemize}

\section{Background}

\begin{figure*}[h]
    \centering
    \includegraphics[width=\linewidth]{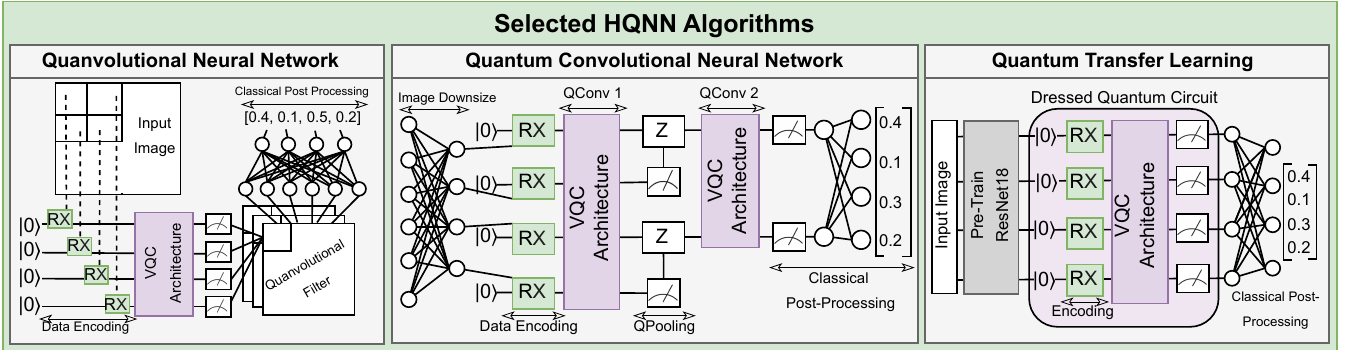}
    \caption{Architecture overview of Sslected HQNN algorithms. Each model utilizes a classical fully connected layer to transform quantum circuit measurement into classification probabilities. In the QCNN, classical convolutional and pooling layers are used for image downsizing to match the qubit count of a circuit.}
    \label{fig:hqnn_soa}
\end{figure*}

We explore three different QNN algorithms amongst the pool of hybrid algorithms, namely, Quanvolutional Neural Networks~\cite{QuanNN:2019}, Quantum Convolutional Neural Networks~\cite{QCNN:2019}, and Quantum Transfer Learning~\cite{QTL:2020}. An overview of the selected algorithms is shown in \Cref{fig:hqnn_soa}. A brief description of the key features of each algorithm and their implementations is presented in the following paragraphs.

\subsection{Quanvolutional Neural Network (QuanNN)}

The Quanvolutional Neural Network (QuanNN) represents an innovative hybrid quantum-classical architecture introduced in~\cite{QuanNN:2019}, which mimics the functionality of classical Convolutional Neural Networks (CNNs) by using quantum computing principles. This model incorporates a new type of transformation layer known as the quanvolutional layer, similar to classical convolutional layers. The layer is composed of multiple quanvolutional filters that process spatially-local subsections of an input tensor, extracting critical features for classification. Each filter is a single circuit that can be customized with specific parameters such as encoding method, type of entangling circuit, number of qubits, and average quantum gates per qubit. We selected the QuanNN as one of our benchmark models because it can be generalized to different sizes of tasks. This is achieved through several configurable aspects: specifying an integer number of quanvolutional filters in each layer, stacking multiple quanvolutional layers together, and customizing its circuit architecture. 
The step-by-step functionality of a QuanNN is: 

\begin{enumerate}[leftmargin=*]
    \item Initiate with a single filter \textit{q} applied to subsections \textit{ux} of dataset images.
    \item Utilize an encoding function \textit{e} to convert  \textit{ux} into an initialized state \textit{ix}.
    \item Process \textit{ix} through a quantum circuit to create an output quantum state \textit{ox}.
    \item Convert \textit{ox} back to guarantee uniform outputs, leading to the final state \textit{fx}.
    \item This entire process is denoted as the ``quanvolutional filter transformation''. \mbox{\textit{fx = Q(ux, e, q, d)}}, where \textit{Q} is the quantum filter.
\end{enumerate}

\subsection{Quantum Convolutional Neural Networks (QCNN)}

Similar to the QuanNN, the QCNN is a QNN algorithm inspired by CNNs. Introduced in~\cite{QCNN:2019}, the QCNN differs from the QuanNN by fully implementing quantum-based convolutional and pooling layers. We selected this architecture for our study as it represents one of the advanced models in QNNs, mirroring the state-of-the-art in classical image recognition. Notably, the QCNN, as presented in~\cite{QCNN:2019}, utilizes only $O(log(N))$ variational parameters for input sizes of $N$ qubits, facilitating its efficient training and practical implementation on NISQ devices.

The QCNN structure comprises an input encoding circuit layer, convolutional circuit layer, pooling circuit layer, and circuit measurement layer. All layers except the last one feature parametric quantum gates. The QCNN is considered an HQNN algorithm because it incorporates classical optimization techniques to adjust the parameterized gate weights. In the quantum convolutional and pooling layers, the interactions among quantum bits efficiently extract features from the input data based on the types of gates used and their arrangement within each layer. However, given the limitations of current NISQ devices, the QCNN can only handle a small number of qubits, limiting its scalability. Thus, it necessitates the use of classical layers to reduce large inputs to suitable sizes for the quantum layers. In our implementation, we utilize single classical convolution and pooling layers to downsize inputs efficiently without losing critical features.

\subsection{Quantum Transfer Learning}

The Quantum Transfer Learning (QTL) is an algorithm inspired by the classical transfer learning model, introduced in~\cite{QTL:2020}. In this algorithm, the knowledge gained from classical neural networks is transferred to quantum neural networks. This process includes basic pre-processing and post-processing of the input and output data with classical layers placed at both the beginning and the end of the quantum neural network. Such a setup results in what is termed a ``dressed quantum circuit'', as in \Cref{eq:Transfer_Learning}. 

\begin{equation}
\hat{Q} = L_{n_{pixel} \rightarrow n_{qubits}} \circ Q \circ L_{n_{qubits} \rightarrow n_{classes}}
\label{eq:Transfer_Learning}
\end{equation}

This hybrid approach is particularly advantageous in the NISQ era, allowing for the optimal pre-processing of high-dimensional data, such as images, using state-of-the-art classical networks. In our model, we employ the pre-trained ResNet-18 model, which is augmented with a VQC serving as the final layer. 
Overview of the QTL Process: 

\begin{enumerate}
    \item Begin with network $A$, which has already been trained on dataset $DA$ for a specific task $TA$. 
    \item Modify network $A$ by removing some of its final layers, thereby creating a truncated version known as network $A'$, which will function as a feature extractor.
    \item Attach a new trainable network, $B$, to the end of $A'$. 
    \item Keep the weights of network $A'$ fixed and concentrate on training network $B$ with a new dataset, $DB$, and for a new target task, $TB$.
\end{enumerate}

\subsection{Quantum Noise \& Error Gates}
Quantum noise refers to the unwanted random fluctuations that occur in quantum systems, due to factors such as environmental interference and uncertainty principle. Unlike classical noise, which often involves predictable and less detrimental effects, quantum noise is more detrimental because of its inherent probabilistic nature, derived from the principles of quantum mechanics. Below we explain different types quantum noise that we used in this paper.

\subsubsection{Bit Flip}

A bit flip error in quantum computing is similar to a bit flip in classical computing. In the classical system, a bit flip error refers to the unintended change of state of a bit from 0 to 1, or vice versa. Similarly, in quantum systems, a bit flip error causes a qubit initially in the state $\ket{0}$ to switch to $\ket{1}$, and vice versa. This type of error is known as a Pauli X error on a qubit. It can be represented using the Kraus Matrices in \Cref{eq:bitflip}, where $ p \in [0, 1]$ denotes the probability of occurrence of bit flip (Pauli X) error. 

\begin{equation}
K_0 = \sqrt{1-p} \begin{pmatrix} 1 & 0 \\ 0 & 1 \end{pmatrix}, \quad 
   K_1 = \sqrt{p} \begin{pmatrix} 0 & 1 \\ 1 & 0 \end{pmatrix}
\label{eq:bitflip}
\end{equation}

\subsubsection{Phase Flip}

Unlike a classical bit, a qubit also has a phase that represents a rotation around the Z-axis. A phase flip is a quantum operation that changes the phase of a quantum state while keeping its probability amplitude intact. The operation is represented by a unitary phase flip gate or Pauli Z gate, which, when applied, leaves the $\ket{0}$ state unchanged and multiplies the $\ket{1}$ state by $−1$. 
This operation only shifts the phase of the $\ket{1}$ state by $\pi$ radians, without affecting the probability amplitude of measuring either state. 
The phase flip error gate is mathematically depicted by the Kraus Matrices in \Cref{eq_phaseflip}, with $p \in [0, 1]$ denoting the probability of encountering a phase flip (Pauli Z) error.

\begin{equation}
K_0 = \sqrt{1-p} \begin{pmatrix} 1 & 0 \\ 0 & 1 \end{pmatrix}, \quad 
   K_1 = \sqrt{p} \begin{pmatrix} 1 & 0 \\ 0 & -1 \end{pmatrix}
\label{eq_phaseflip}
\end{equation}

\subsubsection{Phase Damping}

The phase damping is a type of quantum noise that results in the loss of phase information of a quantum state without altering the probability amplitudes. Unlike bit flip and phase flip noises, the phase damping does not switch the qubit's state from $\ket{0}$ to $\ket{1}$ or vice versa. Instead, phase damping degrades quantum coherence by incrementally reducing the off-diagonal elements of the qubit’s density matrix, representing the superposition's decay. This error can be represented by the Kraus Matrices in \Cref{eq_phasedamping}, where $\gamma \in[0, 1]$ is the probability of phase damping occurring.

\begin{equation}
K_0 = \begin{pmatrix} 1 & 0 \\ 0 & \sqrt{1-\gamma} \end{pmatrix}, \quad 
   K_1 = \begin{pmatrix} 0 & 0 \\ 0 & \sqrt{\gamma} \end{pmatrix}
\label{eq_phasedamping}
\end{equation}

\subsubsection{Amplitude Damping}
The amplitude damping represents a type of quantum noise that occurs when a qubit transitions from an excited state $\ket{1}$ to a ground state $\ket{0}$ due to the loss of energy. This form of error often occurs when a quantum system interacts with an external environment, leading to a gradual energy dissipation over time. Unlike the phase damping, which only modifies the relative phase between the states of a qubit without prompting a transition between states, the amplitude damping involves the probability of a qubit in the excited state $\ket{1}$ decaying to the ground state $\ket{0}$. Amplitude damping errors can be represented using Kraus Matrices in \Cref{eq:ampdamping}, with $\gamma \in [0, 1]$ indicating the probability of amplitude damping.

\begin{equation}
K_0 = \begin{pmatrix} 1 & 0 \\ 0 & \sqrt{1-\gamma} \end{pmatrix}, \quad 
   K_1 = \begin{pmatrix} 0 & \sqrt{\gamma} \\ 0 & 0 \end{pmatrix}
\label{eq:ampdamping}
\end{equation}

\subsubsection{Depolarization Channel}

The depolarizing channel is a noise model in quantum computing that describes a process wherein qubits lose their quantum information to the environment, without favoring any particular basis, Unlike specific errors that only affect certain aspects of a qubit, such as bit flips or phase flips, depolarizing errors are non-selective forms of noise that can randomize a qubit's state to any point on the Bloch sphere with a given probability. 

When a depolarizing error occurs, it can be thought of as the qubit state being replaced with a completely mixed state $\frac{I}{2}$ (where $I$ is the identity matrix) with some probability $p$. This process reduces the purity of the quantum state, effectively ``smearing'' its representation on the Bloch sphere toward the center, corresponding to the maximally mixed state. As $p$ increases, the state becomes more mixed, losing quantum information and coherence. Mathematically, a depolarizing channel is represented with the Kraus Matrices in \Cref{eq:depchannel}, where $p \in [0, 1]$ is the depolarization probability and is distributed evenly across the application of all Pauli operations.

\begin{equation}
K_0 = \sqrt{1-p} \begin{pmatrix} 1 & 0 \\ 0 & 1 \end{pmatrix}, \quad 
   K_1 = \sqrt{\frac{p}{3}} \begin{pmatrix} 0 & 1 \\ 1 & 0 \end{pmatrix}, \quad K_2 = \sqrt{\frac{p}{3}} \begin{pmatrix} 0 & -i \\ 1 & 0 \end{pmatrix}, \quad 
   K_3 = \sqrt{\frac{p}{3}} \begin{pmatrix} 1 & 0 \\ 0 & -1 \end{pmatrix}
\label{eq:depchannel}
\end{equation}

\section{Methodology}
\begin{figure*}[ht]
    \centering
    \includegraphics[width=\linewidth]{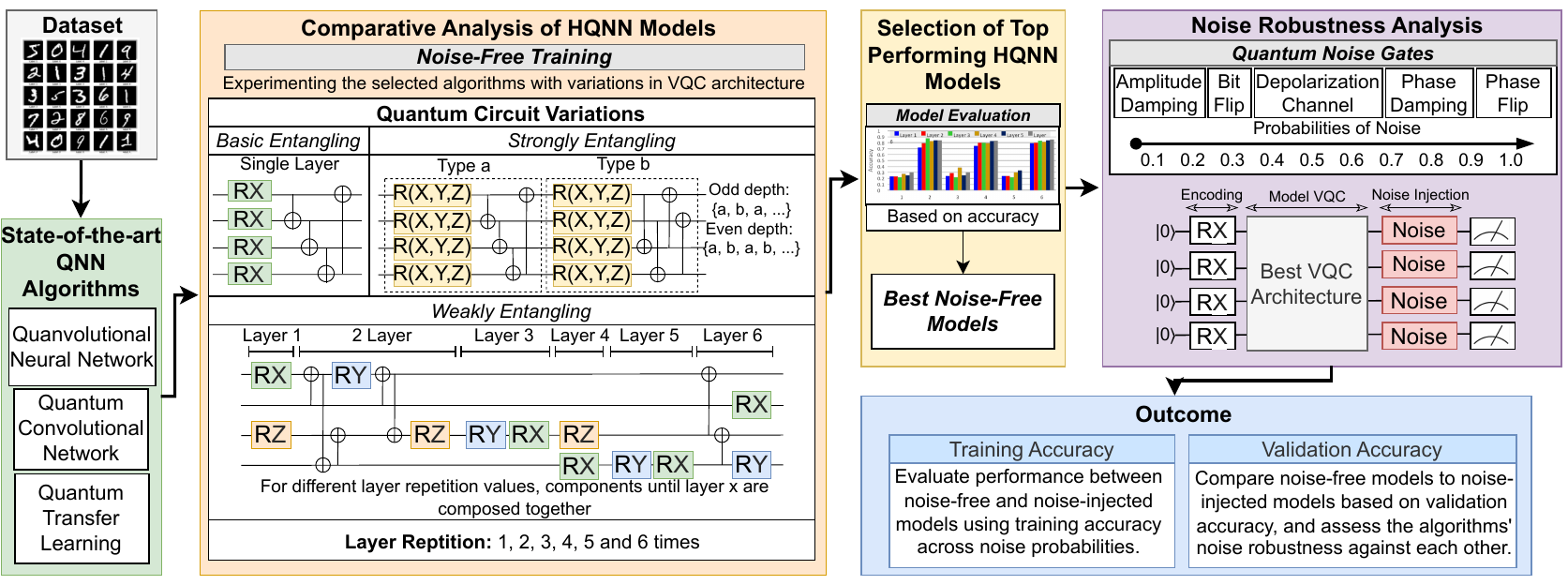}
    \caption{Our methodology. A comprehensive comparative analysis of three different variants of HQNNs is performed with different configurations of quantum layers mainly differing in degree of entanglement, rotation gates, and number of layers (depth of quantum layers). The odd and even depth in a strongly entangling configuration denotes how the layer is repeated when the number of layers are increased. Based on the obtained results the best-performing models with corresponding best configurations are shortlisted which then undergo training under the influence of different types of quantum errors/noise across a wide range of probabilities of each noise type. The comparative analysis of ideal and noisy scenarios is then performed to test the noise-robustness of different HQNN variants. The evaluation metrics used for all the experiments are training and validation accuracy.}
    \label{fig:methodology}
\end{figure*}

In this paper, we analyze three different HQNN algorithms to assess how the circuit architecture impacts performance in an ideal environment without noise. We then explore the effects of different device noise by evaluating the performance of the most effective models in noisy conditions. Our analysis is divided into two main sections: noise-free and noise robustness analysis. \Cref{fig:methodology} provides a detailed overview of our methodology.

\subsection{Dataset Specifications}

In this analysis, we employ a subset of the MNIST dataset~\cite{MNIST}, chosen for its simplicity and effectiveness in facilitating the training process for image classification tasks. The primary focus of our paper is to analyze the effect of noise on model performance; thus, the straightforward nature of the MNIST dataset is adequate to meet the objectives of our study.

The MNIST dataset typically consists of ten classes. However, for our experimental framework, we have restricted our dataset to the first four classes—[0, 1, 2, 3], to ensure that the number of classes directly corresponds to the number of qubits in the quantum circuits utilized in our models. 

\subsection{Selected State-of-the-art HQNN Algorithms}

Our analysis aims to assess the performance of the HQNN models on image classification tasks, particularly examining the impact of noise on their effectiveness. To this end, we selected the three most widely used state-of-the-art HQNN models for our study: the Quanvolutional Neural Network, the Quantum Convolutional Neural Network (QCNN), and Quantum Transfer Learning. 

\subsection{Comparative Analysis of HQNN Models}
Each algorithm we analyze utilizes different classical network and quantum circuit architectures that employ quantum mechanics uniquely. To evaluate the influence of these architectural differences on their performance, we conduct experiments using different VQC variations under noise-free conditions. This approach allows us to evaluate their performance in ideal conditions, providing insights into which circuit configurations are most effective for specific algorithms. 

\subsubsection{Circuit Architecture Variations}
In our experiment, we alter two key variables within the circuit architecture: the type of entanglement and the number of layers. As detailed in \Cref{fig:methodology}, we employ three distinct types of entangling circuits: Weakly Entangling, Basic Entangling, and Strongly Entangling. These circuits differ in the types of rotation gates used and the placement of CNOT gates.

The layer repetition variable indicates how many times a single-layer circuit is repeated in the overall VQC architecture before measurement. The number of layers is analogous to the circuit's depth; more layers result in increased depth. The Basic Entangling and Strongly Entangling circuits have a uniform architecture for each layer, meaning that with each repetition, the same architecture is repeated, thus deepening the circuit. In contrast, the Weakly Entangling circuit employs a non-uniform architecture, with each layer featuring a different composition of gates, as illustrated in \Cref{fig:methodology}.

\subsection{Selected Models}
Based on the noise-free performance of the HQNN models across various circuit configurations, we select the models within each algorithm that achieve an accuracy outcome of above 80\%. This selection criteria allows us to identify the best-performing models for further noise robustness analysis.

\subsection{Noise Robustness Analysis}
The primary objective of this paper is to examine the performance of different HQNN algorithms under various noise conditions. To achieve this, we train the best-performing HQNN models, as identified in the previous analysis, with different noise gates and probability of occurrence. We introduce the noise-inducing gates at the end of the VQC, just before the measurement, as illustrated in \Cref{fig:methodology}. 

\subsubsection{Quantum Noise Gates}
In our analysis, we utilized five types of noise gates: Bit Flip, Phase Flip, Phase Damping, Amplitude Damping, and Depolarization Channel Noise. These gates were selected due to their frequent occurrence in NISQ devices and their ease of implementation using the PennyLane library. Each noise gate is associated with a probability value, indicating the likelihood that it will act on the circuit and modify the quantum state based on its noisy characteristics. We experimented with probabilities ranging from 0.1 to 1.0, in increments of 0.1, yielding a total of ten probability values for each type of noise gate. Note that the number of shots for all the experiments is fixed to 1024, and we do not consider statistical noise caused by varying the number of shots, given the probabilistic nature of quantum computing. 
This experimental design allows us to analyze how the learning capabilities of the models are impacted by the probability of noise occurrence and to explore any potential correlation between model performance and the likelihood of error in a quantum circuit.

\subsection{Outcome}
To assess the robustness of the HQNN models against noise, our analysis will focus on the following key performance metrics: training and validation accuracies. We will conduct a thorough comparison of these accuracies under two distinct conditions: with and without noise in the quantum circuits. This comparison will help us understand how the introduction of quantum noise in the VQC impacts HQNN performance. Furthermore, it will offer insights into the interplay between different characteristics of noise gates and their impact on the model’s robustness or vulnerability to noise.




\section{Results and Discussions}

\subsection{Comparative Analysis of Different HQNN Models}
\begin{figure*}[ht]
    \centering
    \includegraphics[width=\linewidth]{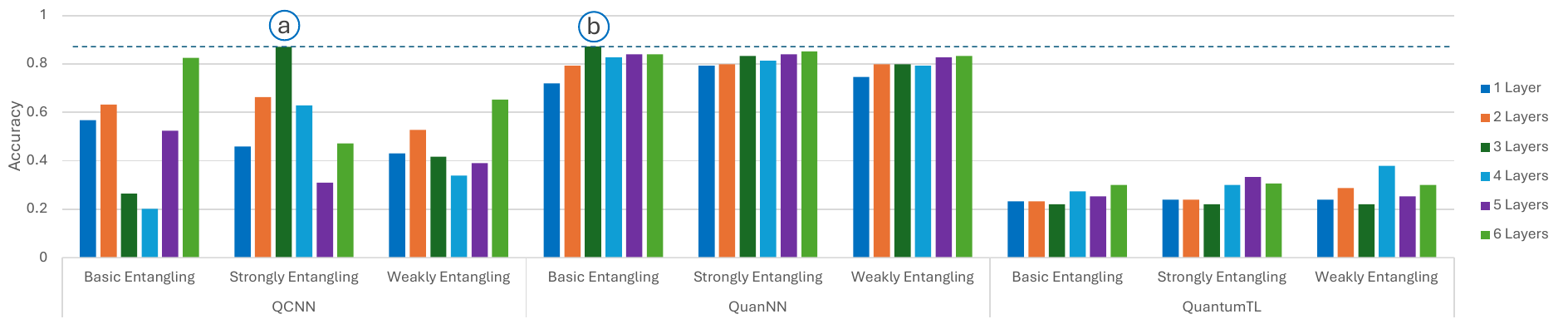}
    \caption{Comparative analysis of QCNN, QuanNN, and QTL in noise-free settings. The HQNN models are tested with different configurations of quantum layers. QuanNN turns out to be the best HQNN variant in terms of both better performance than other variants and also robustness against quantum layer's variations. the second best HQNN model is QCNN, which is sensitive to the degree of entanglement in the underlying quantum layers, i.e., the more the better. QTL consistently performs poorly among the three HQNN variants regardless of the quantum layer configurtation. }
    \label{fig:comparative_analysis_results}
\end{figure*}

We first present the comparative analysis of various HQNN models while assessing their performance on a multiclass classification task.
Our analysis includes experimenting with three different HQNN architectures: QuanNN, QCNN, and QTL, each incorporating a different VQC design as their quantum layer(s) (See Fig.~\ref{fig:methodology}). 
These VQCs differ primarily in the types of parameterized quantum gates utilized (RX, RY, RZ) and the degree of entanglement (categorized as basic, strong, and weak). Furthermore, for all the HQNNs used, the $4$-qubit VQC is consistent across all the experiments. However, we systematically increase the depth of the VQCs from $1$ to $6$ layers to explore the impact of VQC depth on the performance of the HQNN models. The results of our comparative analysis are presented in Fig.~\ref{fig:comparative_analysis_results}. Below, we separately discuss the performance of each HQNN model.

\paragraph{QCNN}

In our analysis of QCNN, we observed that QCNNs exhibited inconsistent performance across different configurations, with no clear trend emerging as the number of layers increased or as modifications were made to the underlying VQC design. This inconsistency can largely be attributed to the QCNN architecture as originally proposed in~\cite{QCNN:2019}, which incorporates mid-network measurements to facilitate pooling operations by tracing out some of the measurement results. Given the probabilistic nature of quantum computation, these mid-network measurements can significantly amplify statistical measurement noise, leading to erratic performance outcomes.

Our results further indicate that the performance of QCNNs is heavily influenced by the degree of entanglement in the underlying the quantum layers.
Specifically, when the quantum layers utilize weak entanglement, QCNNs exhibit suboptimal performance compared to configurations with basic or strong entanglement, regardless of the layer depth. Moreover, to achieve relatively enhanced performance with weakly entangled quantum layers, an increase in layer depth is generally necessary for QCNNs.

On the other hand, QCNNs configurations with basic to strong entanglement demonstrate marked improvements in performance compared to weakly entangled quantum layers. For quantum layers with strong entanglement, moderate depth is found to be optimal, yielding superior performance. 
However, in scenarios where the entanglement is basic, a higher layer depth is required to attain improved performance outcomes. 
This dependency on both the degree of entanglement and the depth of quantum layers underscores the complex interplay of factors that influence the efficacy of QCNN architectures.


\paragraph{QuanNN}

Unlike QCNNs, the QuanNN model demonstrates notably consistent performance across varying depths of layers and different degrees of entanglement. In our analysis, the degree of entanglement appears to have a minimal impact on the performance of QuanNN, with negligible differences observed across the entanglement categories used in this study, i.e., basic, strong, and weak. Additionally, there is a general trend of slight improvement in performance as the depth of the underlying quantum layers increases. This trend suggests that QuanNN benefits from the enhanced expressibility offered by deeper quantum layers, indicating a robustness in its architecture that allows it to maintain performance irrespective of the entanglement degree. This consistent behavior makes QuanNN a potentially more reliable choice in applications where stable performance is critical.


\paragraph{QTL}
The QTL models consistently exhibit inferior performance when compared to both QCNN and QuanNN architectures. Our investigation reveals that the pretrained classical weights from the ResNet18 model do not adapt well to integration with quantum layers. Regardless of the quantum layer depth or the degree of entanglement, QTL models achieve approximately 20\% accuracy, which is significantly lower than the best-case accuracies of over 80\% observed for QCNN and QuanNN. This substantial discrepancy in performance suggests that the QTL architecture may require further optimization or a different approach to effectively leverage quantum enhancements while integrating with pretrained classical network parameters.

In light of the above results, the QTL models underperform significantly compared to QCNN and QuanNN. Consequently, our subsequent analysis on noise robustness will focus exclusively on QCNN and QuanNN models. For these architectures, Table~\ref{tab:best_configs} summarizes the optimal configurations based on our comparative performance evaluation. These configurations will be the only ones considered in the forthcoming noise robustness analysis, ensuring that our investigation concentrates on the most effective setups for each model.

\begin{table}[ht]
    \centering
    \caption{Optimal Configurations for Best Performing Models }
    \begin{tabular}{c|c|c}
    
    \hline
       Model  & VQC Design & \# of Layers \\ 
       \hline
       QCNN   & Strongly Entangling  &$3$ \\

       \hline
        QuanNN & Basic Entangling  & $3$ \\
       \hline
    \end{tabular}
    
    \label{tab:best_configs}
\end{table}


\subsection{Noise Robustness Analysis of QCNN}
We now present the robustness analysis of QCNNs against different types of quantum noise and errors used in this paper. 
This analysis involves deliberately introducing different quantum noises into the quantum layers of both the QuanNN and QCNN models
By systematically injecting these noises at varying probabilities, we aim to assess and compare the resilience of each architecture under different noisy conditions. This approach allows us to understand how these models withstand operational disruptions caused by quantum errors, providing crucial insights into the practical robustness of quantum-enhanced machine learning models.

\begin{figure*}[ht]
    \centering
    \includegraphics[width=\linewidth]{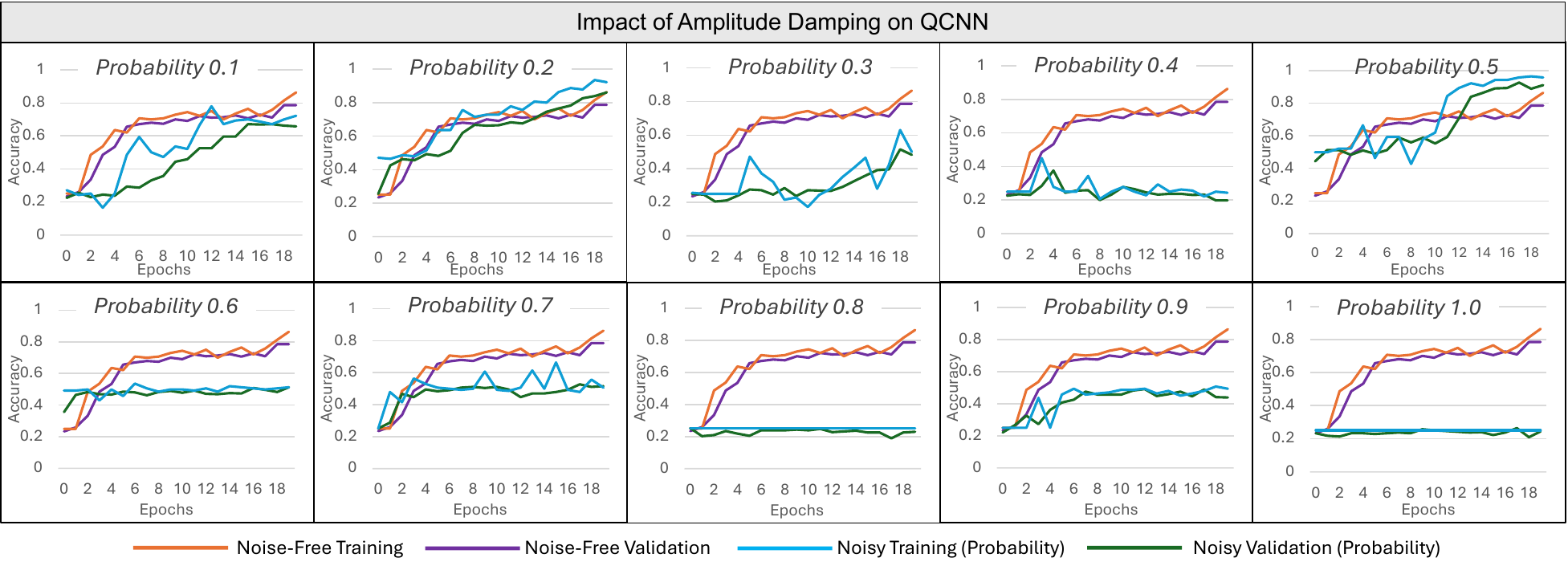}
    \caption{Comparison of QCNN performance in noise free and under Amplitude Damping quantum noise with different probabilities.}
    \label{fig:qcnn_ad}
\end{figure*}

\paragraph{QCNN robustness against Amplitude Damping Noise}
The training and validation results of QCNN under both noisy (amplitude damping) and noise-free scenarios are presented in Fig.~\ref{fig:qcnn_ad}. Our findings reveal that at lower noise probabilities, specifically $0.1$ and $0.2$, QCNNs demonstrate significant resilience to noise, maintaining performance levels comparable to those in an ideal, noise-free setting. However, as the noise probability increases to $0.3$, QCNNs still exhibit some learning potential, but at a significantly slower rate. 
At a noise probability of $0.5$, QCNNs show an unexpected improvement, achieving performance that surpasses the ideal scenario. This anomaly suggests that under certain conditions, the noise might play a constructive role, potentially acting as a form of noise-enhanced learning.

Nonetheless, with further increases in noise probability beyond $0.5$, the QCNNs succumb completely to the adverse effects of the noise, resulting in no learning. 
This highlights the critical impact of noise levels on the operational viability of QCNNs, underscoring the need for optimal noise management strategies in quantum computing applications.


\paragraph{QCNN robustness against Bit Flip Noise}

\begin{figure*}[ht]
    \centering
    \includegraphics[width=\linewidth]{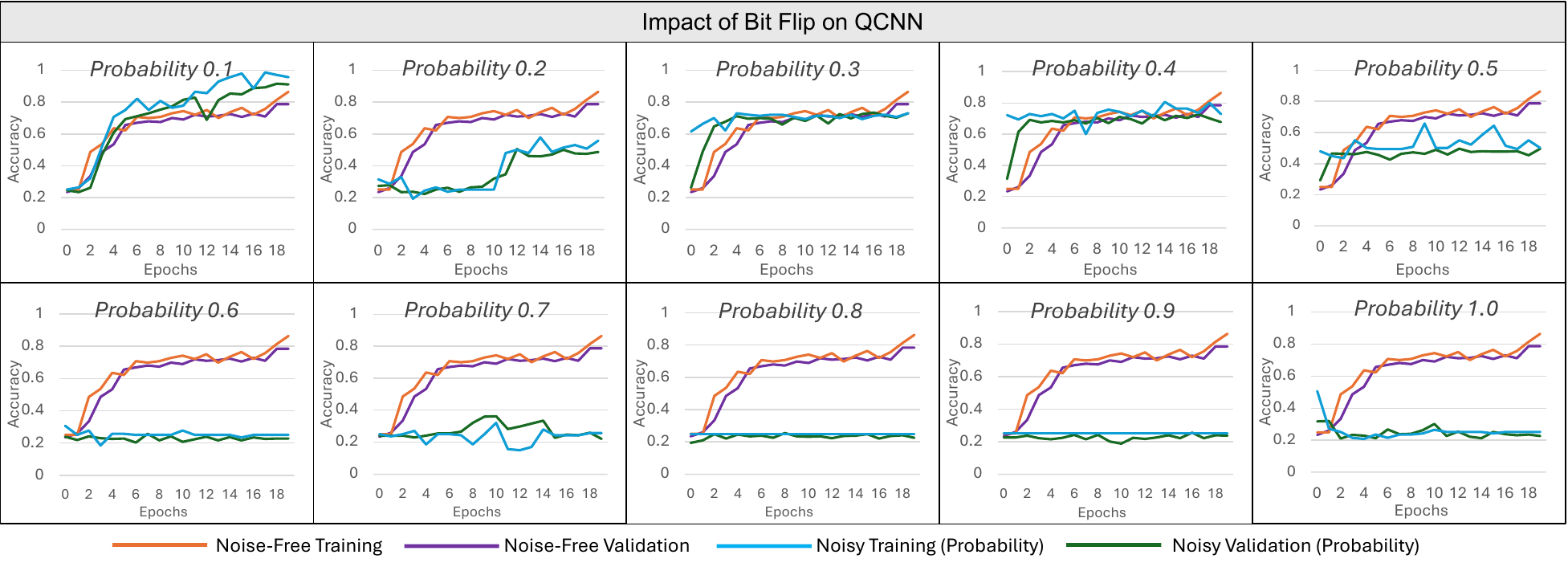}
    \caption{Comparison of QCNN performance in noise free and under Bit Flip quantum noise with different probabilities.}
    \label{fig:qcnn_bf}
\end{figure*}

The comparative analysis of QCNN performance under ideal conditions and when subjected to Bit Flip noise is illustrated in Fig.~\ref{fig:qcnn_bf}. Our results show that at relatively lower noise probabilities (i.e., $\leq 0.4$), QCNNs adapt well to noise patterns, exhibiting great robustness against Bit Flip noise and achieving performance levels comparable to those observed in the noise-free setting. Notably, at very low noise probabilities, QCNNs demonstrate an ability to effectively learn these noise patterns, resulting in training accuracy that exceeds that of the ideal case. However, it is important to note that while training accuracy increases, the generalization (validation accuracy) remains consistent with the ideal setting. This phenomenon suggests that learning from noise can potentially mislead practitioners about the actual performance improvement, as it does not enhance the model's generalization capabilities.
At higher noise probabilities (i.e., $\geq 0.5$), the QCNN succumbs to the detrimental effects of the noise and fails to learn. This highlights a critical threshold in Bit Flip noise tolerance, beyond which the QCNN's performance degrades significantly, indicating the limits of noise adaptability in QCNN architectures.

\paragraph{QCNN robustness against Depolarization Channel Noise}

\begin{figure*}[ht]
    \centering
    \includegraphics[width=\linewidth]{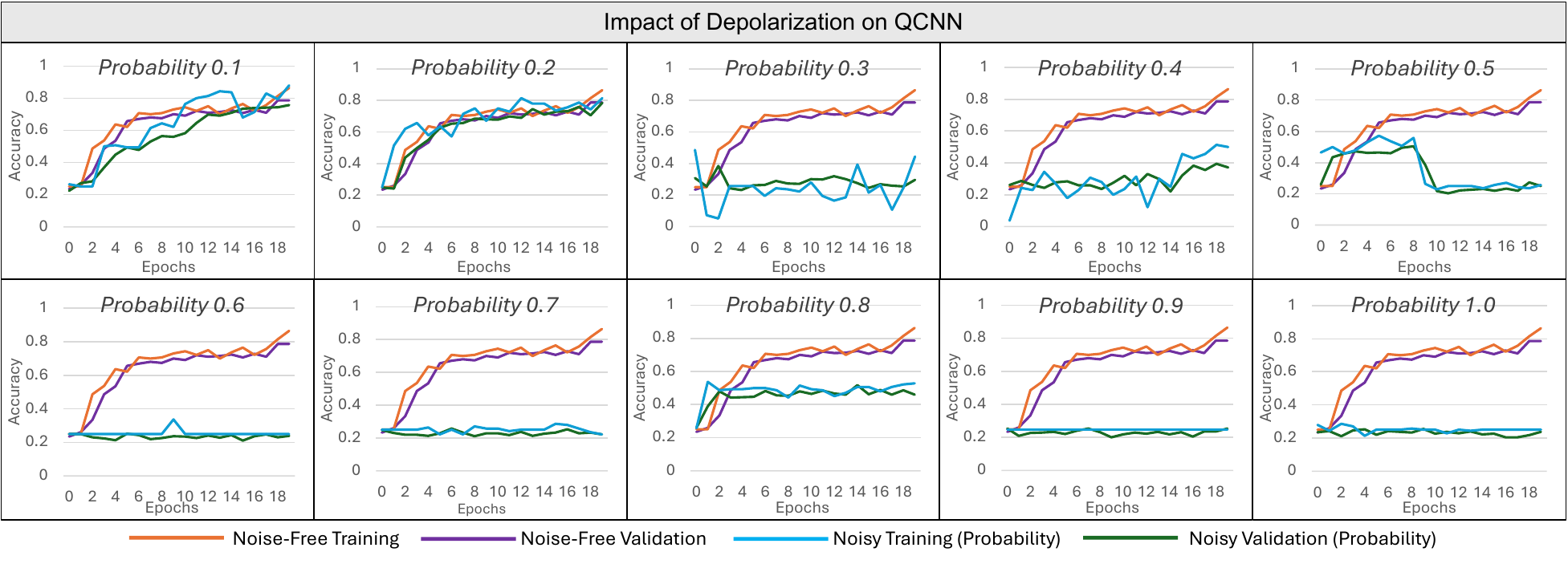}
    \caption{Comparison of QCNN performance in noise free and under Depolarizing Channel quantum noise with different probabilities.}
    \label{fig:qcnn_dep}
\end{figure*}

The performance of QCNNs under both ideal conditions and when subjected to depolarization noise is presented in Fig.~\ref{fig:qcnn_dep}. Our results indicate that at lower probabilities of depolarization channel noise, QCNNs effectively adapt to the noise patterns, demonstrating a robustness that results in performance levels equivalent to those observed in the noise-free setting. This suggests that at these lower noise intensities, QCNNs can tolerate and possibly even utilize minor noise as a feature rather than a detriment, maintaining overall performance.
However, as the noise probability increases beyond $0.2$, the QCNNs succumb to the adverse effects of the noise, showing a significant decline in learning ability. 

This performance degradation can be attributed to the fact that as the noise level increases, the error rates associated with quantum gates may exceed the threshold that the network can compensate for, leading to an accumulation of errors that degrades the quantum state's integrity.
Additionally, higher levels of depolarization noise can lead to a rapid loss of quantum coherence, essential for maintaining the superposition and entanglement necessary for quantum computation. This loss critically hinders the QCNN's ability to process information quantum mechanically.

\begin{figure*}[ht]
    \centering
    \includegraphics[width=\linewidth]{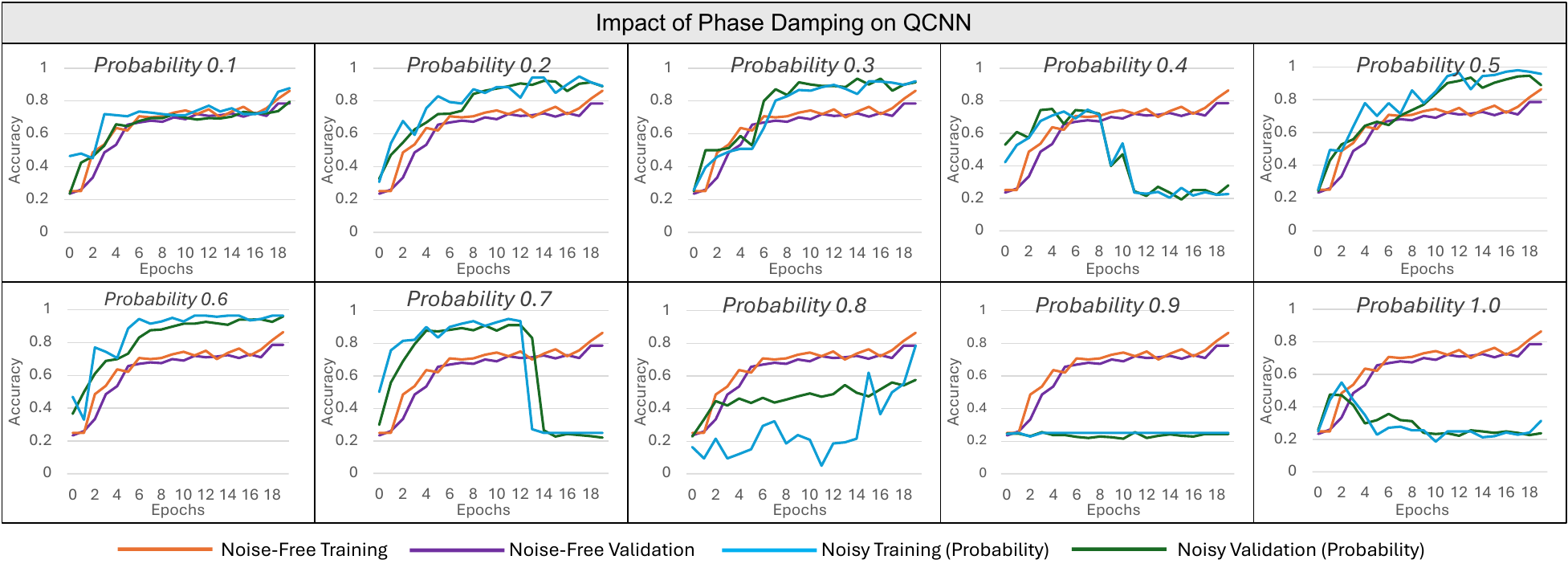}
    \caption{Comparison of QCNN performance in noise free and under Phase Damping quantum noise with different probabilities.}
    \label{fig:qcnn_pd}
\end{figure*}

\paragraph{QCNN robustness against Phase Damping Noise}
The performance of QCNNs under both ideal conditions and when subjected to phase damping noise is presented in Fig.~\ref{fig:qcnn_pd}. 
This analysis diverges from the patterns observed with other types of quantum noise, such as amplitude damping, bit flip, and depolarization channels. Remarkably, QCNNs exhibit considerable robustness to phase flip noise across a broad range of probabilities.

Specifically, at probabilities $\leq 0.3$, QCNNs not only successfully adapt to noise patterns but also perform significantly better than in the ideal, noise-free case. This suggests that phase damping noise, at these levels, might be leveraged as a beneficial factor, possibly aiding in the reduction of other types of quantum errors or enhancing certain quantum state properties that improve computational outcomes.

However, at noise probabilities of $0.4$ and $0.7$, while QCNNs initially adapt to the noise patterns, continued training leads to detrimental effects from the noise. This observation implies that shorter training durations might be beneficial under these specific noise conditions to avoid the negative impact of prolonged exposure to high noise levels.

Interestingly, at probabilities of $0.5$ and $0.6$, QCNNs not only cope with the noise but actually exceed the performance of the ideal scenario. This superior performance under noisy conditions suggests that QCNNs can exploit certain characteristics of phase damping noise to enhance computational accuracy and robustness.

\paragraph{QCNN robustness against Phase Flip Noise}

\begin{figure*}[ht]
    \centering
    \includegraphics[width=\linewidth]{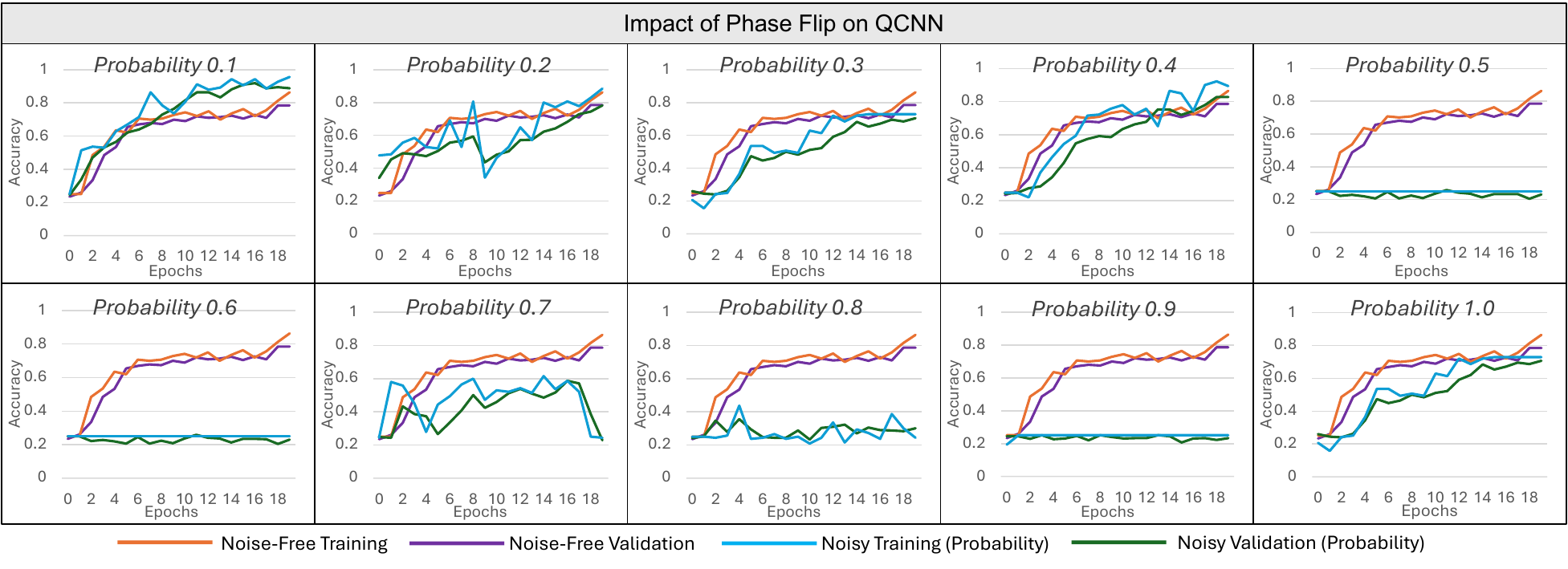}
    \caption{Comparison of QCNN performance in noise free and under Phase Flip quantum noise with different probabilities.}
    \label{fig:qcnn_pf}
\end{figure*}

The performance of QCNNs under both ideal conditions and when subjected to phase flip noise is presented in \Cref{fig:qcnn_pf}. Our results demonstrate that QCNNs adapt well to noise patterns at lower probabilities of phase flip noise (i.e., $\leq 0.4$), maintaining performance levels comparable to those in a noise-free setting.

However, at higher noise probabilities, specifically at 0.5 and 0.6, the QCNN succumbs to the detrimental effects of the noise and fails to learn. Interestingly, at a noise probability of 0.7, the model initially shows resilience, but continued training eventually leads to performance degradation due to the noise. At probabilities of 0.8 and 0.9, the model fails to learn altogether.

Most notably, at a noise probability of 1.0, the model demonstrates remarkable resilience, effectively learning and becoming tolerant to a high intensity of noise. This inconsistent performance across different noise intensities suggests that the QCNN architecture's ability to learn and adapt to noise patterns varies, illustrating its potential resilience and robustness under specific conditions.

\subsection{Noise Robustness Analysis of QuanNN}

\paragraph{QuanNN Robustness against Amplitude Damping Noise}

\begin{figure*}[ht]
    \centering
    \includegraphics[width=\linewidth]{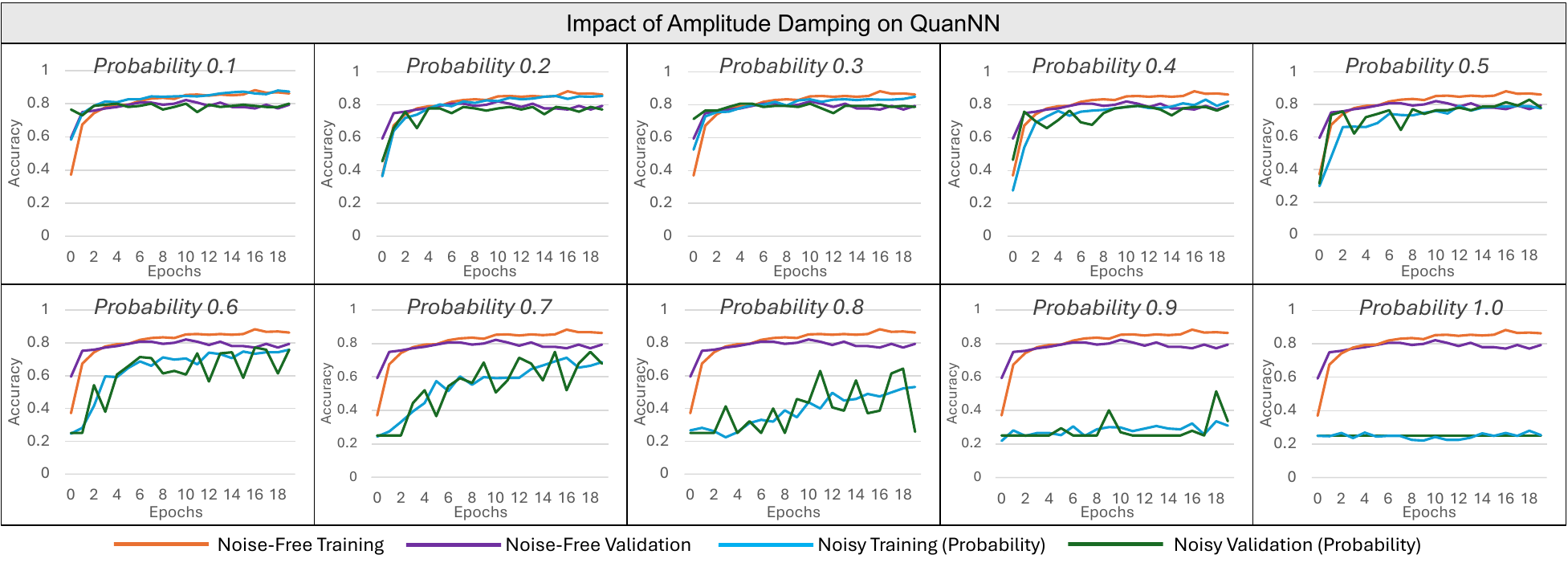}
    \caption{Comparison of QuanNN performance in noise free and under Amplitude quantum noise with different probabilities.}
    \label{fig:quan_ad}
\end{figure*}

Based on the training and validation results of QuanNN under both noise-free conditions and when subjected to amplitude damping noise, as depicted in \Cref{fig:quan_ad}, we observe a consistent trend in the learning capability of QuanNN. Our results show that QuanNN maintains the same learning capacity as in noise-free conditions up to a noise probability of 0.5, demonstrating that the model can retain good learning capabilities for low noise levels.

However, as the probability of noise occurrence increases by increments of 0.1, the learning capability of the models gradually diminishes. At probabilities of 0.6 and 0.7, the performance of QuanNN declines slightly as compared to the performance in a noise-free setting. With further increases in noise, particularly beyond a probability of 0.7, the performance drops drastically, eventually leading to negligible learning at a noise probability of 1.0. This trend underscores the need for improvements in the circuit's ability to handle higher levels of noise during the learning process. 

\begin{figure*}[ht]
    \centering
    \includegraphics[width=\linewidth]{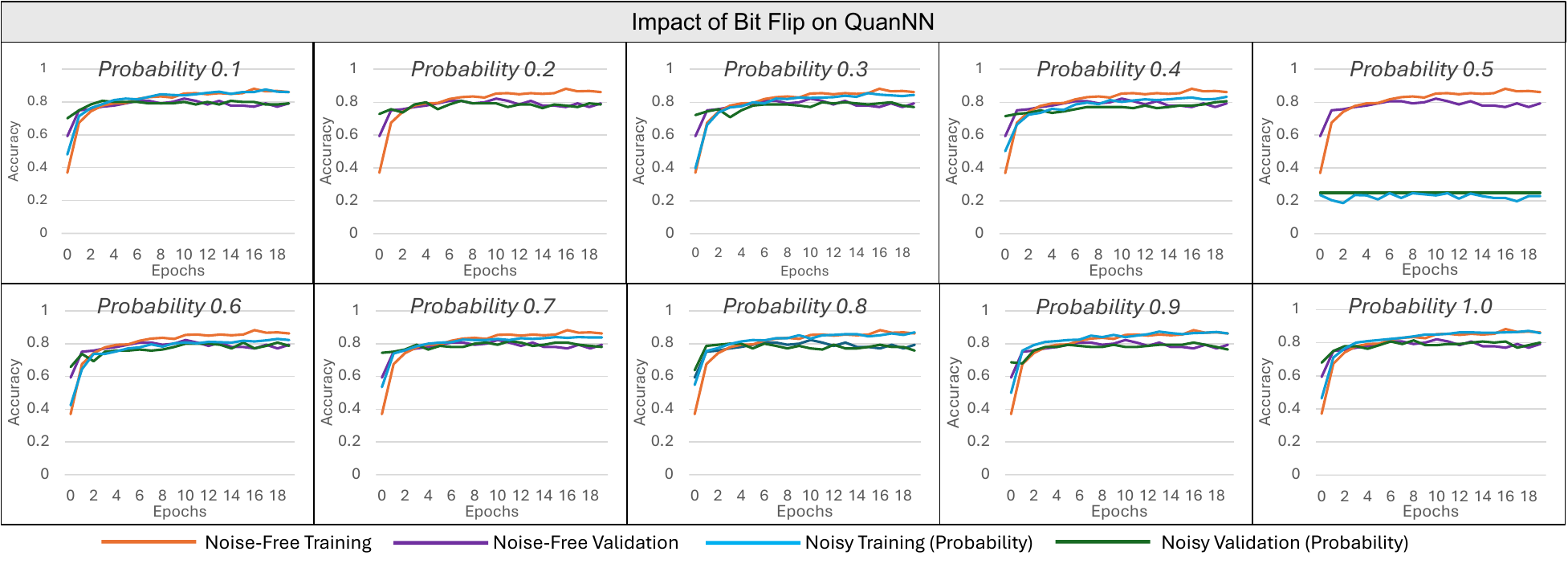}
    \caption{Comparison of QuanNN performance in noise free and under Bit Flip quantum noise with different probabilities.}
    \label{fig:quan_bf}
\end{figure*}

\paragraph{QuanNN Robustness against Bit Flip Noise}
The performance of QuanNN under both ideal conditions and when subjected to bit flip noise is illustrated in Fig.~\ref{fig:quan_bf}. Unlike the performance of QuanNN under amplitude-damping noise, we can observe that the validation accuracy remains consistent with its noise-free performance across most probabilities. This indicates that QuanNN is highly resilient to the effects of bit flip noise at all tested probabilities, with the exception of a significant drop at a noise probability of 0.5.

At a probability of 0.5, the model succumbs completely to the adverse effects of bit flip noise, hindering the learning process. This anomalous behavior can be attributed to the model’s inability to adapt to high noise levels. However, subsequent improvements in performance indicate that the model can adapt to increasing noise characteristics during the learning process. Overall, these results suggest that the architecture of QuanNN is highly robust against bit flip noise.

\paragraph{QuanNN Robustness against Depolarization Channel Noise}

\begin{figure*}[ht]
    \centering
    \includegraphics[width=\linewidth]{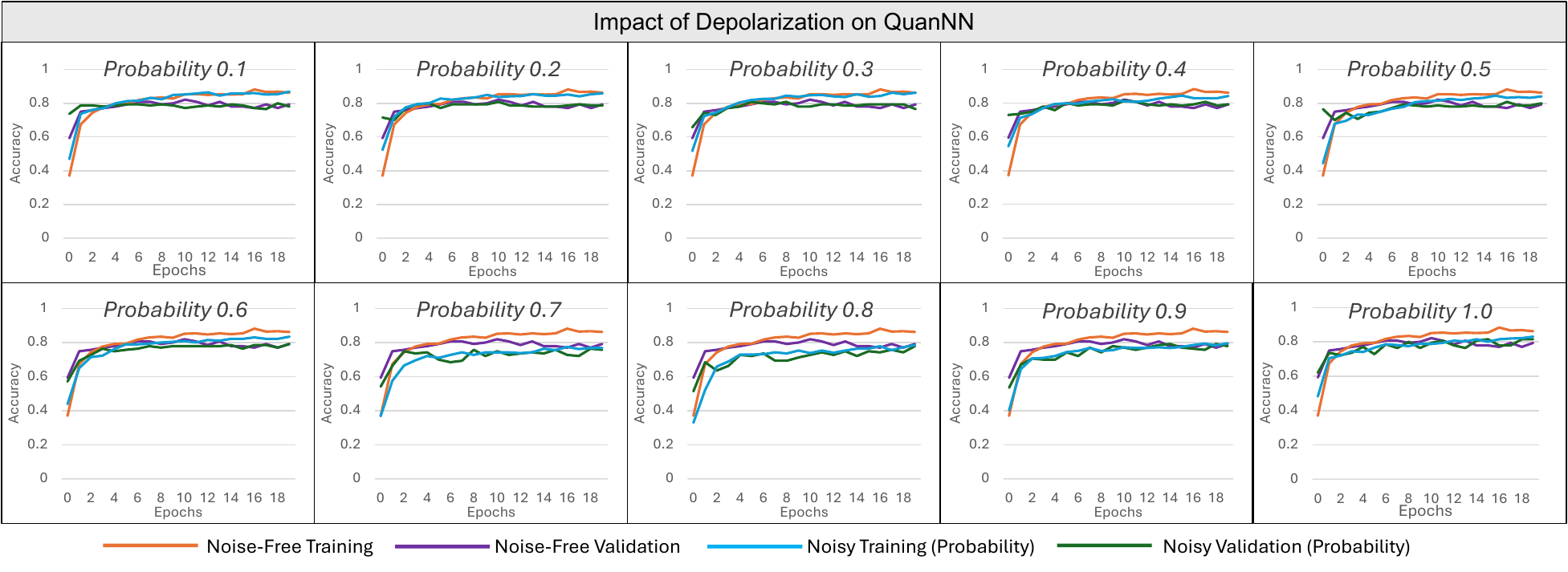}
    \caption{Comparison of QuanNN performance in noise free and under Depolarizing Channel quantum noise with different probabilities.}
    \label{fig:quan_dep}
\end{figure*}

The performance of QuanNN under both ideal conditions and when subjected to depolarization noise is presented in Fig.~\ref{fig:quan_dep}. Unlike the performance of QCNN under depolarization channel noise, QuanNN demonstrates a consistent resilience to all probabilities of depolarization noise, demonstrating a robustness that results in performance levels equivalent to those observed in the noise-free setting.

Although there is a slight decrease in performance at probabilities of 0.7 and 0.8, the model's performance increases from 0.9 onwards, matching that of the noise-free conditions. Typically, higher levels of depolarization channel noise can lead to a rapid loss of quantum coherence, which may hinder the model's learning capabilities~\cite{kashif2024investigating}. However, the results showing QuanNN's performance under increasing probabilities of depolarization noise suggest that its network and circuit architecture makes it very robust against such adverse effects of depolarization noise.

\paragraph{QuanNN Robustness against Phase Damping Noise}

\begin{figure*}[ht]
    \centering
    \includegraphics[width=\linewidth]{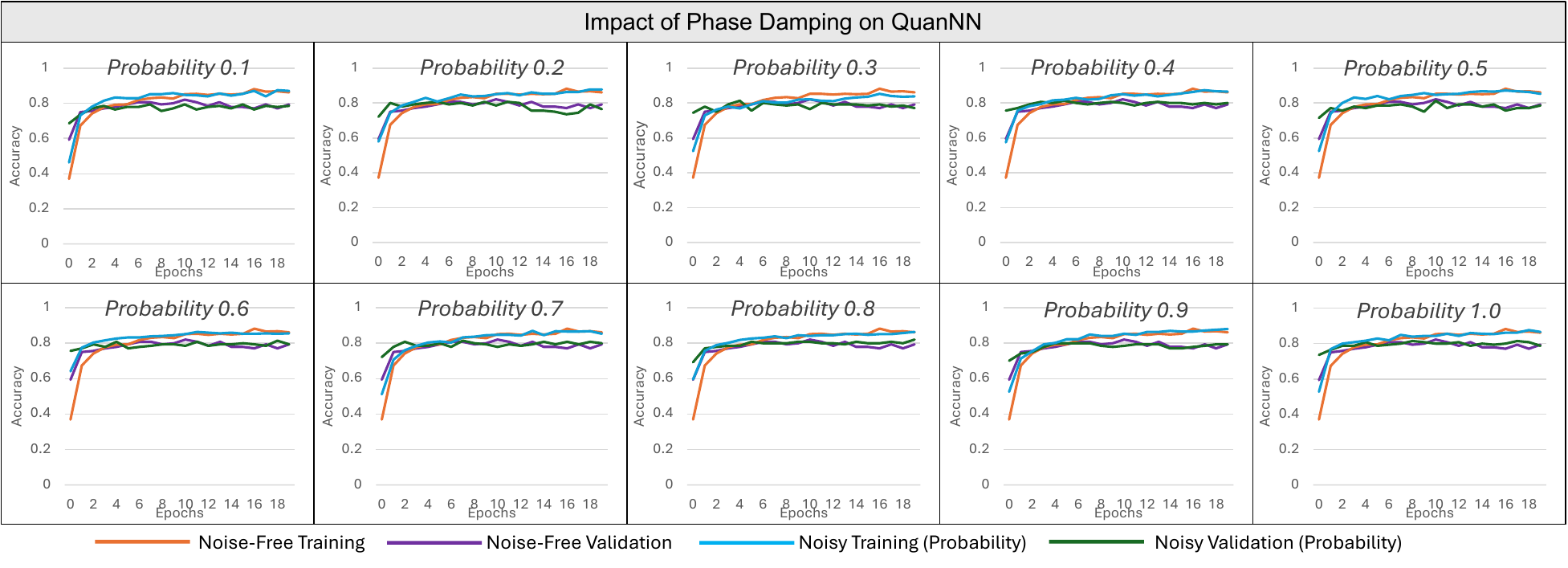}
    \caption{Comparison of QuanNN performance in noise free and under Phase Damping quantum noise with different probabilities.}
    \label{fig:quan_pd}
\end{figure*}

Based on the training and validation results of QuanNN under both noise-free conditions and when subjected to phase damping noise, as depicted in \Cref{fig:quan_pd}, we can observe that QuanNN exhibits considerable robustness to phase damping noise across all probabilities. This resilience is significant as phase damping typically represents a challenge by causing the loss of quantum information without energy loss, potentially degrading the performance. 
Moreover, unlike other forms of quantum noise such as amplitude damping and bit flip, the phase damping noise shows a less adverse effect on QuanNN. This resilient behavior suggests QuanNN architecture’s inherent ability to effectively adapt to and mitigate the effects of phase damping. Additionally, when compared to the performance of the QCNN model under the same conditions, QuanNN exhibits less variability in performance across different noise intensities. This consistency illustrates that the QuanNN architecture is more robust against various noise intensities, highlighting its capability for reliable performance in applications where maintaining phase coherence is essential.

\paragraph{QuanNN Robustness against Phase Flip Noise}

\begin{figure*}[ht]
    \centering
    \includegraphics[width=\linewidth]{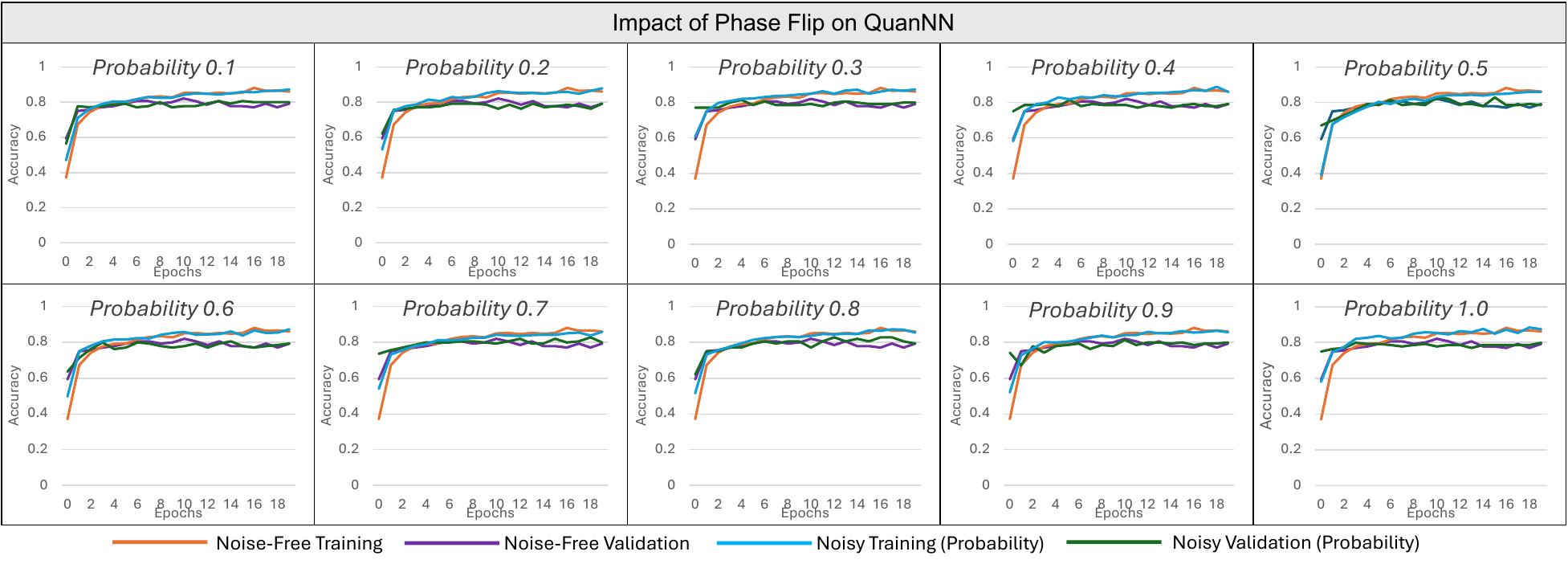}
    \caption{Comparison of QuanNN performance in noise free and under Phase Flip quantum noise with different probabilities.}
    \label{fig:quan_pf}
\end{figure*}

The performance of QuanNN under both ideal conditions and when subjected to phase flip noise is presented in \Cref{fig:quan_pf}. Similar to the performance of QuanNN under phase damping noise, QuanNN also exhibits strong performance across all intensities of phase flip noise. At each probability of noise occurrence, the validation results from the noise-induced conditions overlap with those of the noise-free model. The robust performance of QuanNN demonstrates that QuanNN architecture is highly resilient against noises that affect the phase information of the quantum state.

\section{Conclusion}

In this paper, we have conducted a comprehensive analysis of hybrid quantum neural networks (HQNNs) within the framework of noisy intermediate-scale quantum (NISQ) devices, focusing on their applicability and robustness against variety of quantum erros/noise, in image classification tasks.
Our extensive comparative analysis of various HQNN models, including Quantum Convolution Neural Network (QCNN), Quanvolutional Neural Network (QuanNN), and Quantum Transfer Learning (QTL), has highlighted significant disparities in performance dependent on the entangling structures, layer counts (depth of underlying quantum layers), and their specific configurations within the networks. The comparative analysis is aimed to find the short list the models which perform best for the defined task. Subsequently, a comprehensive noise-robustness analysis of shortlisted models was conducted.

Through a rigorous evaluation of the above-mentioned variants of HQNNs under both ideal and noisy conditions, we have demonstrated that the performance of HQNNs can significantly vary with the introduction of quantum noise.  Our findings reveal that QuanNN consistently outperforms other models across different types of quantum noise, thus highlighting QuanNN as a more robust option in NISQ devices. This superior performance underlines the critical role of model and architecture selection based on the specific quantum noise characteristics of the operational environment.  

Our work contributes to the ongoing development of quantum computing applications by providing detailed insights into the design and optimization of HQNNs for NISQ devices. Additionally, this work offers practical guidance for selecting HQNN architectures that are optimized for resilience to quantum noise, thereby enhancing the reliability and efficacy of quantum-enhanced machine learning solutions.

To summarize, our work not only advances our understanding of the capabilities and limitations of current HQNN models but also sets a foundation for future explorations aimed at harnessing the full potential of quantum computing in real-world applications.

\section*{Acknowledgments}

This work was supported in part by the NYUAD Center for Quantum and Topological Systems (CQTS), funded by Tamkeen under the NYUAD Research Institute grant CG008.




\bibliography{main}







\end{document}